\documentclass[fleqn,twoside]{article}
\usepackage{espcrc2}

\usepackage{graphicx}
\usepackage[figuresright]{rotating}

\newcommand{\AmS}{{\protect\the\textfont2
  A\kern-.1667em\lower.5ex\hbox{M}\kern-.125emS}}

\title{QCD Down Under: Building Bridges}

\author{M.R. Pennington\address{Institute for Particle Physics Phenomenology, University of Durham, Durham DH1 3LE, U.K.}}

\begin{document}

\begin{abstract}
The strong coupling regime of QCD is responsible for 99\% of hadronic phenomena. Though considerable progress has been made in solving QCD in this non-perturbative region, we nevertheless have to rely on a disparate range of models and approximations. If we are to gain an understanding of the underlying physics and not just have numerical answers from computing \lq\lq black'' boxes, we must build bridges between the parameter space where models and approximations are valid to the regime describing experiment, and between the different modellings of strong dynamics. 
We describe here how the Schwinger-Dyson/Bethe-Salpeter approach provides just such a bridge, linking physics, the lattice and experiment.

\vspace{1pc}
\end{abstract}

\maketitle

\section{LIGHT QUARK MASSES}

This meeting which starts today in the beautiful Barossa valley is about \lq\lq building bridges'' between different approaches to QCD, and between these theoretical studies and experiment. I will leave others to discuss work  on the light cone and on flux tubes. Consequently, this talk will serve as an introduction to that by Peter Tandy~\cite{tandy04}, which follows. He will provide the details. I will give the broad brush description and introduce the basic ideas.

Most discussions of QCD begin with the 
perturbative calculations which are so successful in describing a whole range of hadronic phenomena from deep inelastic scattering to beauty production at the Tevatron. These calculations work because they study very short distance interactions well inside the femto-universe when the vacuum appears to be essentially empty, like that of QED. However, what makes QCD so much more fascinating is the fact that over the distance of a fermi it becomes strong. Strong physics is responsible for confinement, for chiral symmetry breaking and for the whole spectrum of hadrons.
This is what I want to review in this talk. 
The vacuum is then far from empty. It is not just a sea of ${\overline q}q$ pairs and a cloud of gluons, the effects of which can be perturbatively computed, but so strong is the interaction that the quarks, antiquarks and gluons form condensates that change the very nature of the vacuum.
The effect of this is illustrated by the way the mass of an up or down quark changes as it propagates over the size of a hadron. As pictured in Fig.~1, at short distances the $u/d$ quark has a very small {\it current} mass but over bigger distances, the size of a fermi, the medium through which it travels generates a {\it constituent} mass, which is about a third of the mass of a proton.

\begin{figure}[hb]
\vspace{-3mm}
\hbox to\hsize{\hss
\includegraphics[width=6.5cm]{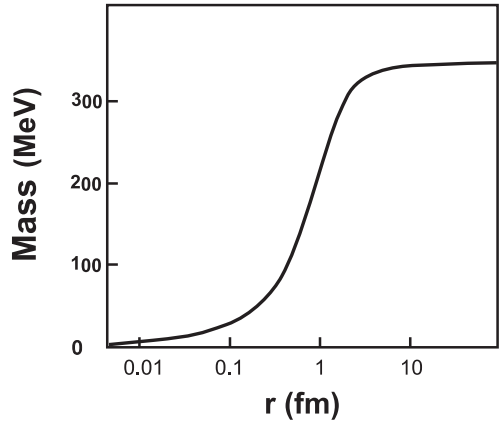}
\hss} 
\vspace{-6mm}
\caption{Schematic illustration of how the mass of an up and down quark depends on the distance $r$ over which it propagates. The quark becomes {\it increasingly} dressed from 'current to constituent' as $r$ increases. }
\end{figure}
   Like all strong physics problems,
this requires calculations beyond the perturbative.

Strong physics problems dominate the world of light quarks, (ups and downs).
 The fact that these quarks are light is crucial to the real world.    
  Indeed hadron jets in $e^+e^-$ annihilation remember the direction
and spin of the initially produced quark and antiquark at a centre-of-mass energy as low as  a few GeV entirely because  most hadronisation requires the creation of ${\overline q}q$ pairs with current masses of just a few MeV. These in turn create pions, which are so much lighter than any other hadron. If the lightest quark were 1.5 GeV in mass, like the charm quark, the world would be quite different and we would have little hint in hadron interactions of the underlying quark dynamics till very much higher energies.

\begin{figure}[t]
\hbox to\hsize{\hss
\includegraphics[width=7.5cm]{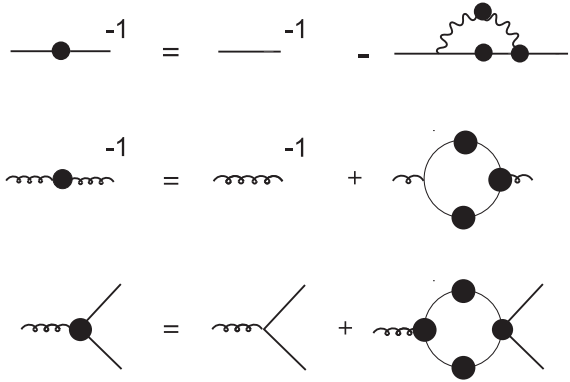}
\hss} 
\vspace{-6mm}
\caption{Schwinger-Dyson equations for the 2 and 3-point functions of QED. The quantities marked with a solid dot are dressed. The solid lines represent fermions, the wiggly lines photons. }
\vspace{-3mm}
\end{figure}  
But how does this picture, shown in Fig.~1, emerge from QCD. To answer this, let us consider more generally when can masses be dynamically generated in a gauge theory --- can the bare mass be zero and yet the particle have mass~\cite{miransky,qed4}? It is well known that in perturbation theory, the mass of any particle is proportional to its bare mass at every order. So if the bare mass is zero, the dressed mass remains zero. Consequently, mass generation must be a strong physics problem. A possible approach to studying this would be to consider the field theory on the lattice. However, massless particles do not fit on a finite size lattice, so calculations have to be performed with non-zero mass and the massless result obtained by extrapolation. Such an extrapolation has little to do with the lattice computations, but has to be calculated in some other way, as we will discuss again later. This means we have to treat massless gauge theories in the continuum. The field equations of the theory are the Schwinger-Dyson equations, Fig.~2. These are genuinely non-perturbative and are in the form of a set of nested integral equations. 

To study the question of dynamical mass generation at its simplest, let us consider the electron in QED.
 The dressed fermion propagator is determined from the bare fermion propagator by the dressed fermion, the dressed photon and the full fermion-photon interaction, as in Fig.~2. The dressed photon is in turn determined by dressed fermions and the full fermion-photon vertex. One can imagine solving the coupled electron and photon 2-point equations, if   we know the full fermion-photon interaction, but this of course satisfies
its own Schwinger-Dyson equation  (Fig.~2) that relates it to the 2 and 3-point functions we have discussed and a new 4-fermion interaction, which in turn is related to 2, 3, 4 and 5-point functions. We have an infinite system of coupled equations.
\begin{figure}[h]
\hbox to\hsize{\hss
\includegraphics[width=7.5cm]{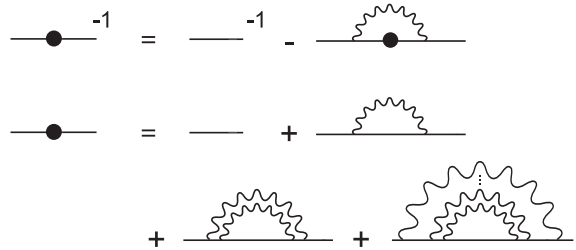}
\hss} 
\vspace{-3mm}
\caption{Schwinger-Dyson equation for the inverse fermion propagator in an 
approximation, where only the fermion 2-point function is dressed. This corresponds to the summation of the  {\it rainbow} graphs shown in the lower equation.}
\end{figure}
It appears that to find the 2-point function we need to know all $n$-point functions, unless we can find a way to truncate this system. The only self-consistent truncation we know is perturbation theory, but this is quite inappropriate for the problem of mass generation. So let us instead make a simple, but brutal truncation of the Schwinger-Dyson equations.

 Let us treat the full fermion-photon interaction as bare, viz. $\gamma^{\mu}$, and quench the photon propagator in Fig.~2. This gives the {\it rainbow} approximation shown in Fig.~3. Then the coupling $\alpha$ is a constant, $\alpha_0$, that does not run. Consequently, the only momentum scale is provided by the ultraviolet cut-off $\kappa$, which would be replaced by some physical scale on renormalization.
\begin{figure}[t]
\hbox to\hsize{\hss
\includegraphics[width=7.5cm]{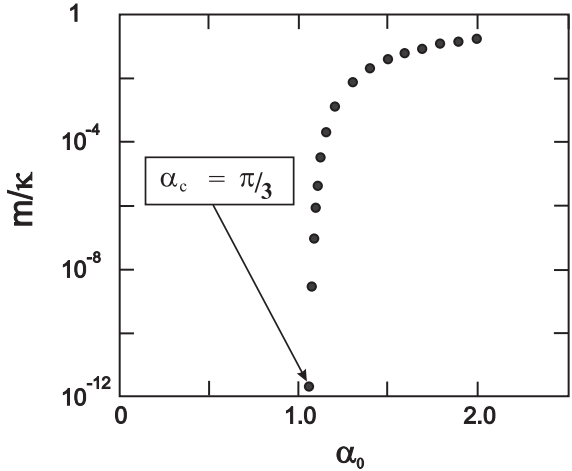}
\hss} 
\vspace{-6mm}
\caption{Fermion mass defined at Euclidean momenta by ${\cal M}(p^2=m^2)=m$ in units of the ultraviolet cut-off, $\kappa$, as a function of the constant coupling $\alpha_0$, with zero bare mass, showing how a dynamical mass is generated for $\alpha_0 > \alpha_c = \pi/3$.}
\vspace{-6mm}
\end{figure} 
 With this butchery we have a non-perturbative equation for the electron propagator, which couples its wavefunction renormalization and its mass function. In the Landau gauge the wavefunction is unrenormalised and we just have to solve for the electron mass.  An infinite set of equations has been reduced to one. This can then be readily solved~\cite{miransky,qed4}. One finds, as shown in Fig.~4, that with a zero bare mass the dressed electron remains massless until the coupling becomes strong at some critical value, which in this case is $\pi/3$. As the coupling strengthens beyond this value, a larger and larger mass is generated. 
What this means in the world of quenched QED is illustrated by considering  the behaviour of a massless electron in the presence of nuclei of increasing atomic number. The electron would move at the speed of light until the atomic number was greater than 140 and then it would slow up as the strength of the interaction generates a mass.

If one performs this calculation in any other covariant gauge than Landau a mass is generated above some critical value, but this coupling strength depends on the gauge, which is of course unphysical. This is because the rainbow approximation violates the Ward-Green-Takahashi identity (WGTI)~\cite{wgt}. To make the answer gauge independent, one not only has to make the interaction fulfill the WGTI but the fermion equation must be multiplicatively renormalisable too~\cite{curtis}. This requirement essentially determines the projection of the full fermion-photon interaction imposed by the fermion Schwinger-Dyson equation --- at least in quenched QED. We learn that if the coupling is strong enough then masses can be generated and that is all we need to motivate a programme of  study in QCD, where the $u$ and $d$ quarks are so nearly massless, but hadrons have mass!
\begin{figure}[h]
\hbox to\hsize{\hss
\includegraphics[width=7.5cm]{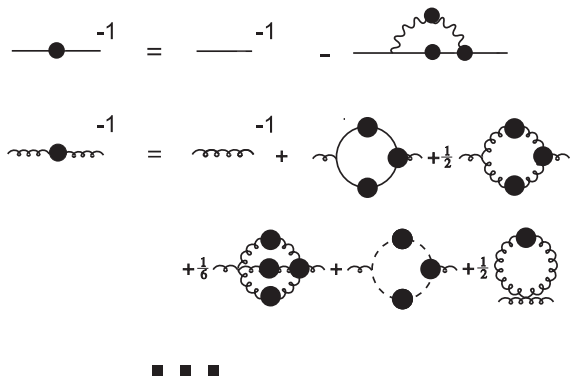}
\hss} 
\vspace{-3.5mm}
\caption{Schwinger-Dyson equations for the quark and gluon propagators in QCD in a covariant gauge. The solid lines are fermions, the wiggly lines gluons and dashed ghosts. The solid dot indicates fully dressed quantities.}
\end{figure} 

QCD is inherently more complicated. There are of course the gluon self-interactions as well as ghosts in covariant gauges, Fig.~5. The first Schwinger-Dyson studies of the infrared coupling in QCD were by Pagels~\cite{pagels}, Mandelstam~\cite{mandelstam}, and
 Bar-Gadda~\cite{bargadda}, \footnote{other pioneering studies like that of Baker, Ball and Zachariasen~\cite{BBZ} were in axial gauges, but they used a truncation that was subsequently shown to be seriously inconsistent~\cite{West}.}
with an extensive numerical investigation by Nick Brown and myself~\cite{brown}.\begin{figure}[t]
\hbox to\hsize{\hss
\includegraphics[width=7.5cm]{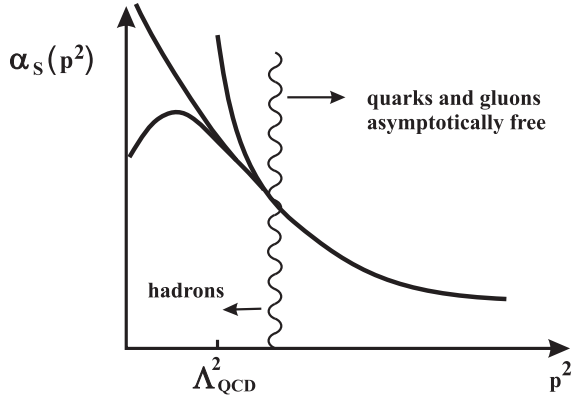}
\hss} 
\vspace{-2mm}
\caption{Coupling in QCD as a function of momentum, $p$. For $p \gg \Lambda_{QCD}$ the running shows {\it asymptotic freedom}, while for $p < \Lambda_{QCD}$
we have different possibilities of confining behaviour. Which is the one required by QCD is an issue of current research.}
\vspace{-2mm}
\end{figure} 
 Here the ghosts were treated perturbatively to produce the right behaviour in the ultraviolet regime. The solutions of the equations of Fig.~5 revealed a gluon propagator that in the infrared became strongly enhanced, as did the strong coupling like the upper curve in Fig.~6. This enhancement was consistent with a linear confining  heavy quark potential and strong enough to produce a dynamical mass for the $u$ and $d$ quarks. This was studied in detail by Maris and Roberts~\cite{marisrbts}, who investigated the quark equation for different bare masses from zero to 5 GeV to span the range from up and down to beauty, Fig.~7.
Provided the combined effect of the quark-gluon interaction and the gluon propagator over the infrared regime is enhanced, a dressed mass of 350-500 MeV is generated, as phenomenology requires --- the scale being set by $\Lambda_{QCD}$. 
The difference between the behaviour with a massless current quark and one of 3-5 MeV is seen to be very small --- except on the logarithmic scale of Fig.~7! 

Dynamical mass generation requires the development of non-zero condensates. Their value is only strictly defined in the chiral limit. The asymptotic behaviour of the mass function is given by
$$
 {\cal M}(p^2)\;=\;m_0\,\left(\ln \frac {p^2}{\Lambda^2}\right)^d\,+\, C\,\frac{\langle \, {\overline q}q\, \rangle}{p^2}\,\left(\ln \frac{p^2}{\Lambda^2}\right)^{-d-1}
$$
where $d$ is the appropriate anomalous dimension and $m_0$ is the current mass.
In the limit $m_0\to 0$, $\langle \; {\overline q}q\; \rangle$ becomes gauge invariant.
By looking at the ultraviolet behaviour in Fig.~7, Langfeld {\it et al.}~\cite{langfeld}  show a ${\overline q}q$ condensate of scale $-(250-300 {\rm MeV})^3$ results, just as the phenomenology of QCD sum-rules~\cite{sumrules} and experiments probing low energy $\pi\pi$ interactions~\cite{leutwyler} would claim.
\begin{figure}[t]
\vspace{-3mm}
\hbox to\hsize{\hss
\includegraphics[width=7.1cm]{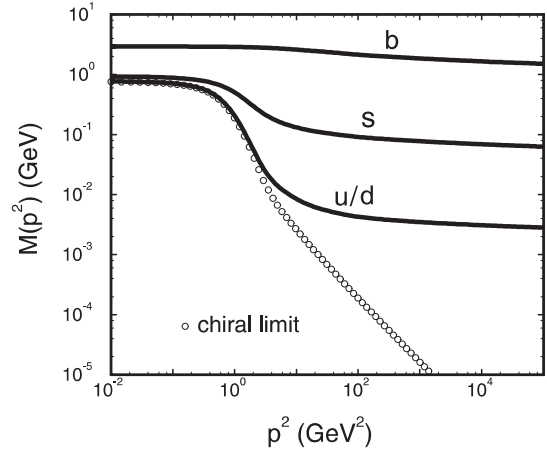}
\hss} 
\vspace{-6mm}
\caption{Euclidean mass function ${\cal M}(p^2)$ as a function of momentum, $p$,
for different current masses corresponding to $b$, $s$ and $u,d$ quarks. The chiral limit of massless current quarks is shown by the circles, from the calculation of Ref.~\cite{marisrbts}.}
\vspace{-2mm}
\end{figure} 

Further studies have however indicated that ghosts might well play a more important role than hitherto expected. This follows from a systematic investigation by the T\"ubingen group~\cite{alkofer} of the coupled ghost, gluon and then quark Schwinger-Dyson equations. With no equivalent of the Slavnov-Taylor identity to guide the construction
 of the needed ghost-gluon vertex these studies have so far been restricted
to the Landau gauge,     where the ghost kernels are known to be simple~\cite{ghostwatson}.
Alkofer and co-workers~\cite{alkofer} found that contrary to previous investigations the gluon does not become enhanced at low momenta, but in fact its dressing function goes to zero.
In contrast, the ghost dressing function is the one that is infrared enhanced. The combination of these means that the effective coupling does still increase at smaller momenta, but with a constant infrared limit, as shown in Fig.~8. Though  the exact value of this limit is known to be dependent on the approximations made, the fact that it is larger than 1
is sufficient to ensure the dynamical chiral symmetry breaking indicated by experiment.

 \begin{figure}[t]
\hbox to\hsize{\hss
\includegraphics[width=7.4cm]{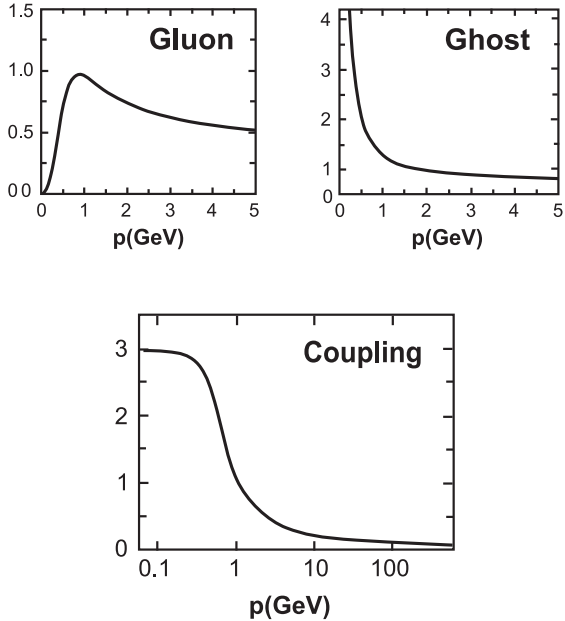}
\hss} 
\vspace{-5mm}
\caption{Momentum dependence of the Landau gauge gluon and ghost dressing functions from  Schwinger-Dyson studies~\cite{alkofer}. These give the {\it effective} quark-gluon coupling shown.}
\vspace{-2mm}
\end{figure}
 This scheme appears to solve one mystery: are gluons confined or confining~\cite{buttner}? If a particle is confined then one does not expect it to have a mass-shell
and so the numerator of its propagator should cancel any potential pole in the denominator. This the T\"ubingen scheme~\cite{alkofer} achieves for the transverse gluons. Nevertheless, we know the effective coupling must be enhanced in the infrared if confinement is to be generated. Here this is produced by the unphysical degrees of freedom of the gluon (and ghost) propagators.
Thus this scheme for the ghost/gluon sector potentially resolves the dilemma and both confines and is confined. However, how this mechanism can actually generate a near linearly confining potential for heavy quark systems is an important issue currently under detailed investigation.

\section{LIGHT HADRON MASSES}

One can input the form of the gluon, ghost and quark functions just discussed into the bound state equations and study the properties of the meson spectrum, particularly light hadrons for which the long range nature of the forces is so critical~\cite{tandy04}.

As emphasised by Roberts and collaborators \cite{craig}, the axial Ward identity ensures that the  ${\overline q}q$ bound state with pseudoscalar quantum numbers is a Goldstone boson with its interactions governed by PCAC. In contrast the bound states with scalar and vector quantum numbers have masses reflecting the mass of the fully dressed (or constituent) quark. The scalar mass is found to be  highly sensitive to the details of the quark scattering kernel~\cite{craig2}
of Fig.~9, 
the vector meson states much less so. Consequently, we will consider these again shortly. Peter Tandy~\cite{tandy04} will flesh out many of these considerations when he describes his extensive study of bound states. The behaviour of the gluon and ghost propagators built into these calculations can be compared with Monte Carlo lattice simulations and are in excellent agreement~\cite{alkofer2,alkofer}. While lattice calculations  can only be performed with
sizeable quark masses, the Schwinger-Dyson/Bethe-Salpeter system being continuum equations can be computed in the massless limit.

\begin{figure}[t]
\hbox to\hsize{\hss
\includegraphics[width=7.5cm]{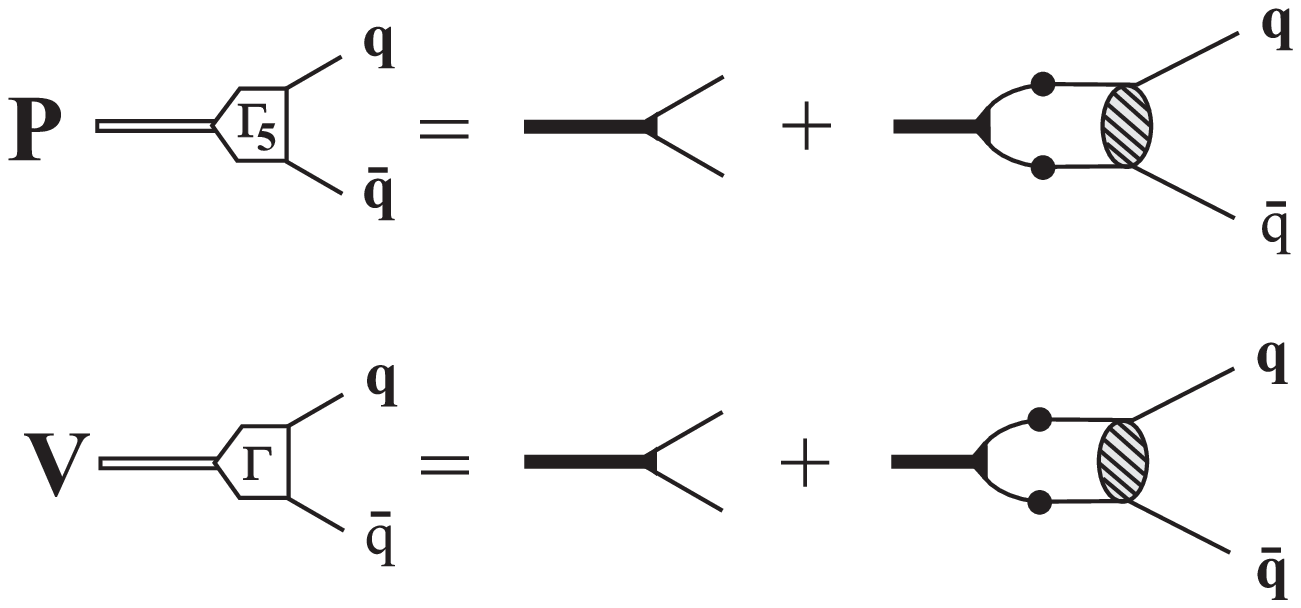}
\hss} 
\vspace{-3mm}
\caption{Bound state equations of Bethe and Salpeter for pseudoscalar and vector mesons. Their solution depends critically on the $q{\overline q}$ scattering kernel indicated in hatched grey.}
\vspace{-3mm}
\end{figure} 
How to continue lattice results to small quark masses has been a major research theme over the last couple of years~\cite{davies}.
For instance, the $\rho$-meson mass depends linearly on the quark mass, with a non-zero value in the chiral limit. For large mass the pion too depends linearly on the quark mass, but   its Goldstone nature means it must have vanishing mass when its quark components are massless. Indeed, if chiral symmetry breaking is dominated by a non-zero ${\overline q}q$-condensate then
the pion mass depends on the square root of the quark mass at small values. Lattice results with quark mass larger than 150 MeV confirm the linear relation between the $\rho$ and $\pi$ masses with barely any deviation. In contrast, chiral perturbation theory ($\chi$PT) requires the square root mass dependence and the presence of logarithms of the pion mass. The key to extrapolation is to know when these set in and how to match these to the behaviour found by lattice calculations.

If $\chi$PT is computed in dimensional regularisation, this is problematic since infrared and ultraviolet behaviours become entangled. Thus as noted by Donoghue {\it et al.}~\cite{donoghue} these effects increase at larger pseudoscalar mass,  but larger masses define the domain of momentum the lattice should treat exactly. The need is to separate the infrared behaviour where chiral dynamics dominates and the lattice treats poorly, from larger momenta where the lattice embodies the correct physics.
Thomas {\it et al.}~\cite{thomas} from Adelaide and Donoghue {\it et al.}~\cite{donoghue} are amongst several sets of authors who have given prescriptions for handling this transition.

Here I want to advertise that the Schwinger-Dyson/Bethe-Salpeter approach naturally encompasses physics at  all momentum scales. Respecting the axial Ward identities ensures that chiral dynamics is included with no need to specify at which momentum this chiral behaviour dominates. The bound state equations determine the dependence of the hadron masses for all quark masses, which would be exact if we knew the ${\overline q}q$ scattering kernel precisely. In principle lattice calculations do the same. In practice they do so only for sizeable quark mass. Indeed, the kaon mass is the lightest pseudoscalar mass that computational limitations  currently allow.
On the lattice all masses are only specified in units of lattice spacing. One can eliminate this dependence entirely by plotting the vector meson mass, for instance, as a function of the pion mass. Results from CP-PACS~\cite{CP-PACS} for different lattice couplings are shown in Fig.~10. 

Using the T\"ubingen modelling of the strong coupling limit of QCD, we can then calculate the very same relationship for all quark masses in the continuum approach.
Following the work of Maris and Tandy~\cite{maristandy}, the dressed gluon propagator can be represented by
$$ \Delta^{\mu\nu}(p)\;=\;C \;\frac{p^2}{\omega^2}\,\exp\left(-\frac{p^2}{\omega^2}\right)\,             \Delta^{\mu\nu}_0(p)$$
where $\Delta_0$ is the bare transverse gluon propagator. The parameter $C$ is related to the strength of the interaction, while $\omega$ corresponds to the momentum at which the gluon dressing peaks shown in Fig.~8~\cite{alkoferwatson}. From the work of Ref.~\cite{maristandy} this is known to be around 0.5 GeV. Using the simple rainbow-ladder approximation to solve the Bethe-Salpeter equation for the $\rho$ and $\pi$-mesons one finds the behaviour shown as the line in Fig.~10.

Optimal agreement~\cite{watson} between Bethe-Salpeter (BS) and CP-PACS results occurs for $\omega = 0.425$ GeV, very close to the expected value. The continuation from a pseudoscalar mass of  500 MeV down to a 140 MeV and the chiral limit is specified.

\begin{figure}[t]
\hbox to\hsize{\hss
\includegraphics[width=7.5cm]{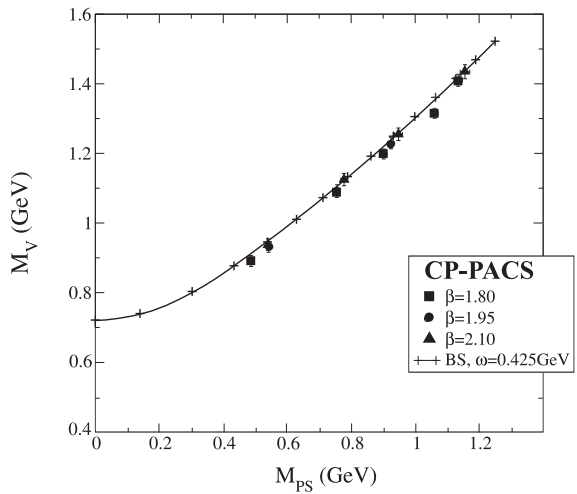}
\hss} 
\caption{Vector meson mass as a function of pseudoscalar meson mass. The data points are from the lattice calculations of CP-PACS~\cite{CP-PACS} as shown in the inset. The curve that connects the crosses is the result of the Bethe-Salpeter calculation using the T\"ubingen modelling of the full quark and gluon functions by Watson {\it et al.}~\cite{watson} The parameter $\omega$ is explained in the text.}
\vspace{-3mm}
\end{figure}

For this continuation one need not estimate where the transition to the dominance of chiral dynamics occurs.
 This is built in. This non-perturbative approach to the continuum knows. 
However, at the present level of approximation this may not yet accord precisely with $\chi$PT. Calculation does show that the pion mass is proportional to the square root of the current quark mass. However, how much of the detailed behaviour embodied in $\chi$PT  encoded in the chiral logs is there in the present solutions of the bound state equations is under study~\cite{watson}.
The calculations compared in Fig.~10 in the continuum and on the lattice are both performed with just two quark flavours. The strange quark is ignored. Of course, at large pion mass (i.e. greater than 450 MeV) the $\rho$ cannot decay. It is stable. The Bethe-Salpeter calculation by Watson {\it et al.}~\cite{watson} does not yet include the contribution from pion loops either. This is known to give a shift of 60 MeV or so in the $\rho$ mass if it is to generate the experimental width, but elements of double counting occur that are under study.
Nevertheless progress surely has been made.  It is not that we have lattice calculations valid for larger quark mass and chiral perturbation theory at small quark mass with educated guesses of how to join them. The Schwinger-Dyson/Bethe-Salpeter approach in the continuum holds out the prospect that we have results applicable at {\it all} quark masses.

In this talk and that that follows by Peter Tandy~\cite{tandy04}, we see that a bridge is being built that not only relates theory to experiment,
but can relate the lattice to the continuum. The Schwinger-Dyson/Bethe-Salpeter connection provides a wide-spanning bridge. The other talks in the week ahead  will doubtless reveal a panorama of interlinking bridges between hadron physics and the fascinating strong coupling regime of QCD down under, that is the subject of this exciting Workshop.


\section*{Acknowledgments}
It is a very great pleasure to thank my fellow organisers of this Workshop for such an interesting, entertaining and productive meeting and the Special Research Centre for the Subatomic Structure of Matter of the University of Adelaide for making this all possible.  I particularly want to express my gratitude to
Ay\c se K{\i}z{\i}lers\"{u}, whose unfailing attention, made this meeting so memorable.
I acknowledge the partial support of the EU-RTN Programme, 
Contract No. HPRN-CT-2002-00311, \lq\lq EURIDICE'' for this work.

\end{document}